\documentclass{article}

\usepackage{arxiv_style}

\usepackage{graphicx} 
\usepackage{amsmath, amsfonts}
\usepackage{hyperref}       
\usepackage{url}            
\usepackage{nicefrac}       
\usepackage{booktabs}       
\usepackage{microtype}      

\title{Scrambler: Mixed Boolean Arithmetic Obfuscation Tool Using E-graph and Equality Expansion}
\author{
 Seoksu Lee \\
  Chungnam National University\\
  Daejeon\\
  Republic of Korea \\
  \texttt{troy.doubles@o.cnu.ac.kr} \\
   \And
 Sangjun An \\
  Chungnam National University\\
  Daejeon\\
  Republic of Korea \\
  \texttt{sangjun0319@gmail.com} \\
  \And
 Eun-Sun Cho \\
  Chungnam National University\\
  Daejeon\\
  Republic of Korea \\
  \texttt{eschough@cnu.ac.kr} \\
}

\begin{document}
\maketitle

\begin{abstract}
We propose Scrambler, and e-graph-based MBA obfuscation tool using Equality Expansion to efficiently generate complex and diverse expressions with equivalence guaranteed by construction.
Experiments show Scrambler improves existing tools in expressiveness and complexity.
\end{abstract}


\section{Instruction}
Program obfuscation transforms a program into a functionally equivalent but less readable form to hinder analysis.
Among various obfuscation techniques, data obfuscation and Mixed Boolean–Arithmetic (MBA) obfuscation are widely used due to their relatively low application cost and high resistance to reverse analysis.
To perform MBA obfuscation, existing obfuscation tools rely on predefined MBA rules and specific matrix-based information (e.g., truth tables and linear algebra-based representations), which limits the diversity of generable MBA expressions.

In this paper, we present Scrambler, an MBA obfuscation tool that leverages e-graphs \cite{egg} and equality expansion to generate diverse obfuscation expressions with guaranteed equivalence.

\section{Background and Related work}
\subsection{MBA Obfuscation}
MBA obfuscation complicates simple expressions by mixing Boolean and arithmetic operators.
For example, $x+y$ can be rewritten as $(x \lor y) + (x \land y)$. 
Such transformations can produce many semantically equivalent but more complex expressions, which are commonly classified as linear MBA expressions, polynomial MBA expressions, or non-polynomial MBA expressions, depending on their complexity.



\subsection{E-graph}
An e-graph\cite{egg} is a graph-type data structure that supports equivalence. 
It adds e-classes to a typical graph data structure to express equivalence, eliminating the need for duplicate representations of similar meanings. 
As shown in the figure \ref{fig:e_graph_example} below, equivalence can be expressed using dotted-line e-classes for cases with the same meaning, allowing multiple expressions to be expressed simultaneously on the graph.

\begin{figure}[h]
    \centering
    \includegraphics[width=0.5\linewidth]{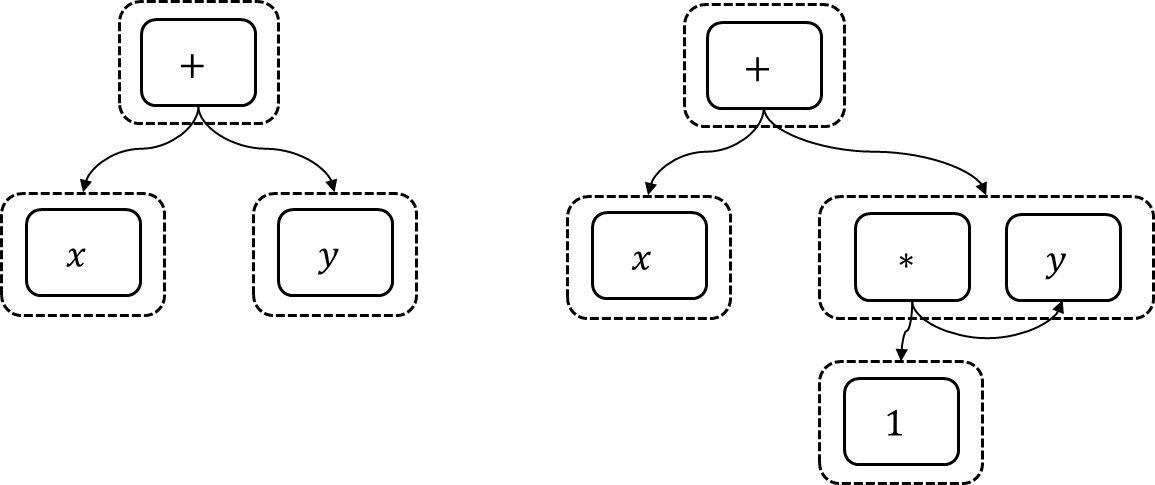}
    \caption{$x + y$ in e-graph representation (left) and after applying the rule $y*1=y$ (right). The * operation and $y$ are labeled with the same e-class, indicating that these two nodes have the same meaning.}
    \label{fig:e_graph_example}
\end{figure}

\subsection{Existing MBA Obfuscation Tools}
The MBA Obfuscator \cite{mbaobfsucator} is a representative MBA obfuscation tool 
capable of generating 
linear, polynomial, and non-polynomial MBA expressions.
However, it requires manual configuration of truth tables and basic expressions depending on the number of variables used, and supports a limited number of operators.

NeuReduce \cite{neureduce} and Loki \cite{loki} also generate MBA expressions for deobfuscation.
NeuReduce generates only linear MBA expressions 
and relies on simple user-provided expressions and truth tables, 
making it unsuitable for complex obfuscation
and incurring high generation overhead.
Loki’s MBA expression generator assigns a \textit{Depth} parameter to each operation, 
allowing iterative expression extension to grow arbitrarily deep for large-scale expression generation.
However, its operator set is more restrictive than that of other tools, and
incurs huge overhead in generation.
As a result, existing MBA obfuscation tools suffer from limitations in operator support, expression types, generation time, and scalability. Moreover, most rely on SMT solvers such as Z3 for equivalence checking, whose performance degrades on large or complex MBA expressions.

\section{Scrambler}
Scrambler is an obfuscation tool proposed in this paper to overcome the limitations of existing MBA obfuscation tools. 
Scrambler utilizes an idea called Equality Expansion, which is an adaptation of the existing Equality Saturation \cite{eqsat} method. 
Equality Saturation is an optimization algorithm that considers all possible cases for all user-provided rules to find the simplest expression. 
This algorithm utilizes the e-graph data structure for this purpose. 
Equality Expansion extends this algorithm to generate complex expressions based on user-provided rules. 
For example, while the termination condition of Equality Saturation is finding the simplest expression, Equality Expansion seeks to find expressions that satisfy the termination condition (e.g. AST size must exceed 1000). 
This method effectively generates complex expressions from simple e-graphs by utilizing the e-graph data structure representing equivalence.

Scrambler is implemented using the Rust egg framework and effectively generates complex MBA expressions by applying the equivalence expansion technique described above.
Scrambler takes as input a rule set, an expression to be obfuscated, and a termination condition, and generates the most complex expression satisfying the conditions.
The equivalence expansion termination condition can be configured with various conditions, including expression size, e-graph rule search conditions, and termination time. Depending on the rules used, various types of MBA expressions, such as linear, polynomial, and non-polynomial, can be generated.
Furthermore, unlike existing obfuscation tools, Scrambler has the advantage of not performing separate verification of the generated expressions. 
This is because, based on the e-graph and equivalence expansion concepts, if the semantic equivalence of the input rules is guaranteed, the generated MBA expression is also guaranteed to be semantically equivalent. 
Therefore, Scrambler can use a wider range of operators and generate more complex expressions faster than existing tools.

\section{Experiment}
We conducted comparative experiments on expressions generated by Scrambler and by existing tools\cite{neureduce_github, loki_github, mbaobfuscator_github}.
Since each tool uses different obfuscation methods, it is difficult to compare expressions generated from the same input expressions.
Therefore, we evaluated MBA-obfuscated expressions using multiple metrics across different input formulas.

(1) \textit{Size}: The size of the generated expression, measured by the number of nodes in its abstract syntax tree (AST).
(2) \textit{Var \#}: The number of variables in the expression.
(3) \textit{Const \#}: The number of constants in the expression.
(4) \textit{Op \#}: The number of operators in the expression.
(5) \textit{MBA Alternation}: A complexity metric for MBA expressions, as defined in \cite{mbaobfsucator}.
(6) \textit{Entropy}: The Shannon entropy of variables and constants, representing expression complexity.

\begin{table}[ht]
\centering
\caption{AVERAGE MBA EXPRESSION METRICS PER OBFUSCATION TOOL}
\label{tab:mba_metrics}
\begin{tabular}{llrrrrrr}
\toprule
Tool & Variant & AST Size & \# of Var & \# of Const & \# of Op & MBA Alternation & Entropy \\ 
\midrule
{NeuReduce} & Original & 3.48 & 1.74 & 0.04 & 1.70 & 0.31 & 1.65 \\
 & Obfuscated & 23.85 & 2.13 & 1.37 & 13.31 & 3.89 & 3.11 \\ 
\midrule
{Loki} & Original & 3.00 & 2.00 & 0.00 & 1.00 & 0.00 & 1.58 \\
 & Obfuscated & 216.17 & 2.00 & 0.00 & 142.82 & 38.26 & 2.66 \\ 
\midrule
MBA & Original & 4.95 & 2.58 & 0.05 & 1.97 & 0.47 & 2.00 \\
Obfuscator & Obfuscated & 235.22 & 2.65 & 8.08 & 136.22 & 31.83 & 3.12 \\ 
\midrule
{Scrambler} & Original & 7.44 & 2.14 & 0.12 & 3.39 & 0.92 & 2.16 \\
 & Obfuscated & 34786.77 & 2.14 & 0.88 & 23373.46 & 6387.58 & 2.59 \\ 
\bottomrule
\end{tabular}
\end{table}

Table \ref{tab:mba_metrics} summarizes the metrics of obfuscated expressions for each tool, where the reported values represent the average of the collected data.
For MBA Obfuscator, we generated three types of obfuscated expressions—linear, polynomial, and non-polynomial—from simple input expressions, resulting in approximately 120 MBA expressions.
For NeuReduce, we generated expressions with two or three variables using the \textit{n\_terms} option set to 5 and 7, and evaluated a total of 150 obfuscated MBA expressions.
For Loki, we set the Depth option to 5 and measured metrics for 5,000 obfuscated MBA expressions covering all supported operators.

Scrambler similarly performed MBA obfuscation on 100 input expressions using 14 rules, with the e-graph traversal option set to 2, the node limit set to 3,000, and the time limit set to 2 seconds.
As shown in Table \ref{tab:mba_metrics}, the results indicate that Scrambler is capable of generating MBA expressions with higher structural complexity compared to existing obfuscation tools.
More robust obfuscated expressions can be obtained by relaxing the equality expansion conditions; however, the extent of such robustness depends on memory availability and the imposed rule constraints.

Because Scrambler applies semantics-preserving rewrite rules, the generated expressions are guaranteed to be semantically equivalent to the original ones, eliminating the need for verification using an SMT solver.
For instance, when no applicable rules are available for a given input expression, no obfuscation is produced.
In contrast, we confirmed that effective obfuscation is achieved when multiple rules are applied.

We experimentally increased the number of rules associated with a single operator and observed an increase in the metric after a certain threshold was exceeded.
  
\section{Conclusion}
In this study, we proposed an equation expansion method based on e-graphs and equation saturation techniques, and implemented Scrambler, an MBA obfuscation tool built upon this approach.
Although a precise comparison with existing MBA obfuscation tools is challenging due to differences in their operational mechanisms, our experimental results indicate that Scrambler can generate MBA expressions with higher structural complexity using approximately 14 rewrite rules and a node expansion limit of 3,000.
Furthermore, by enforcing strict rewrite rules, the semantic equivalence of expanded expressions is inherently preserved by the e-graph framework, eliminating the need for a separate verification step and improving the efficiency of obfuscated expression generation.
Nevertheless, the correctness of the applied rules must be carefully validated, and performance may vary depending on rule selection. Addressing these issues remains an important direction for future work.

\section{Acknowledgment}
This work was supported by the National Research Foundation of Korea (NRF) grant funded by the Korea government (MSIT)(RS-2024-00456953)
  
\bibliographystyle{unsrt}
\bibliography{references}

@inproceedings{mbaobfsucator,
author = {Liu, Binbin and Feng, Weijie and Zheng, Qilong and Li, Jing and Xu, Dongpeng},
title = {Software Obfuscation with Non-Linear Mixed Boolean-Arithmetic Expressions},
year = {2021},
isbn = {978-3-030-86889-5},
publisher = {Springer-Verlag},
address = {Berlin, Heidelberg},
url = {https://doi.org/10.1007/978-3-030-86890-1_16},
doi = {10.1007/978-3-030-86890-1_16},
booktitle = {Information and Communications Security: 23rd International Conference, ICICS 2021, Chongqing, China, November 19-21, 2021, Proceedings, Part I},
pages = {276–292},
numpages = {17},
location = {Chongqing, China}
}

@article{egg,
author = {Willsey, Max and Nandi, Chandrakana and Wang, Yisu Remy and Flatt, Oliver and Tatlock, Zachary and Panchekha, Pavel},
title = {egg: Fast and extensible equality saturation},
year = {2021},
issue_date = {January 2021},
publisher = {Association for Computing Machinery},
address = {New York, NY, USA},
volume = {5},
number = {POPL},
url = {https://doi.org/10.1145/3434304},
doi = {10.1145/3434304},
journal = {Proc. ACM Program. Lang.},
month = jan,
articleno = {23},
numpages = {29}
}

@inproceedings {loki,
author = {Moritz Schloegel and Tim Blazytko and Moritz Contag and Cornelius Aschermann and Julius Basler and Thorsten Holz and Ali Abbasi},
title = {Loki: Hardening Code Obfuscation Against Automated Attacks},
booktitle = {31st USENIX Security Symposium (USENIX Security 22)},
year = {2022},
isbn = {978-1-939133-31-1},
address = {Boston, MA},
pages = {3055--3073},
url = {https://www.usenix.org/conference/usenixsecurity22/presentation/schloegel},
publisher = {USENIX Association},
month = aug
}

@inproceedings{neureduce,
    title = "{N}eu{R}educe: Reducing Mixed {B}oolean-Arithmetic Expressions by Recurrent Neural Network",
    author = "Feng, Weijie  and
      Liu, Binbin  and
      Xu, Dongpeng  and
      Zheng, Qilong  and
      Xu, Yun",
    editor = "Cohn, Trevor  and
      He, Yulan  and
      Liu, Yang",
    booktitle = "Findings of the Association for Computational Linguistics: EMNLP 2020",
    month = nov,
    year = "2020",
    address = "Online",
    publisher = "Association for Computational Linguistics",
    url = "https://aclanthology.org/2020.findings-emnlp.56/",
    doi = "10.18653/v1/2020.findings-emnlp.56",
    pages = "635--644"
}

@inproceedings{eqsat,
  author = {Tate, Ross and Stepp, Michael and Tatlock, Zachary and Lerner, Sorin},
  title = {Equality saturation: a new approach to optimization},
  year = {2009},
  isbn = {9781605583792},
  publisher = {Association for Computing Machinery},
  address = {New York, NY, USA},
  url = {https://doi.org/10.1145/1480881.1480915},
  doi = {10.1145/1480881.1480915},
  booktitle = {Proceedings of the 36th Annual ACM SIGPLAN-SIGACT Symposium on Principles of Programming Languages},
  pages = {264–276},
  numpages = {13},
  location = {Savannah, GA, USA},
  series = {POPL '09}
}

@misc{loki_github,
  key =          {loki},
  year  =        2023,
  title =        "loki",
  howpublished  = {\url{https://github.com/RUB-SysSec/loki}},
  lastaccessed = "Jan 14, 2026",
}

@misc{mbaobfuscator_github,
  key =          {MBA-Obfuscator},
  year  =        2021,
  title =        "MBA-Obfuscator",
  howpublished  = {\url{https://github.com/nhpcc502/MBA-Obfuscator}},
  lastaccessed = "Jan 14, 2026",
}

@misc{neureduce_github,
  key =          {NeuReduce},
  year  =        2020,
  title =        "NeuReduce",
  howpublished  = {\url{https://github.com/fvrmatteo/NeuReduce}},
  lastaccessed = "Jan 22, 2026",
}

\end{document}